\documentclass[prd,onecolumn,notitlepage,longbibliography,preprintnumbers,nofootinbib,onecolumn]{revtex4-1}
\usepackage[utf8]{inputenc}

\usepackage{bm}
\usepackage[colorlinks=true,urlcolor=blue,anchorcolor=blue,citecolor=blue,linkcolor=blue,filecolor=blue,menucolor=blue,pagecolor=blue,linktocpage=true,pdfproducer=medialab,pdfa=true]{hyperref}
\usepackage{amsmath,latexsym,amssymb,mathrsfs,ascmac}
\usepackage{graphicx}

\begin{document}
\preprint{KOBE-COSMO-19-16}
\title{Birth of de Sitter Universe from time crystal }
\author{Daisuke Yoshida}
\email{dyoshida@hawk.kobe-u.ac.jp}
\affiliation{Department of Physics, Kobe University, Kobe 657-8501, Japan}
\author{Jiro Soda}
\email{jiro@phys.sci.kobe-u.ac.jp}
\affiliation{Department of Physics, Kobe University, Kobe 657-8501, Japan}

\begin{abstract}
We show that a simple sub-class of Horndeski theory can describe a time crystal Universe.
 The time crystal Universe can be regarded as a baby Universe nucleated from a flat space, which is mediated by an extension of Giddings-Strominger instanton in a 2-form theory dual to the Horndeski theory. Remarkably, 
when a cosmological constant is included, 
 de Sitter Universe can be created by tunneling from the time crystal Universe.
 It gives rise to a past completion of an inflationary Universe.
\end{abstract}

\maketitle

\section{Introduction}
Inflation has succeeded in explaining current observations of the large scale structure 
of Universe~\cite{Planck:2013jfk,Ade:2013sjv}. 
However, inflation has a past boundary \cite{Borde:2001nh}, which could be a true initial singularity or just a coordinate singularity depending on how much the Universe deviates from the exact de Sitter \cite{Yoshida:2018ndv}, or depending  on the topology of the Universe \cite{Numasawa:2019juw}. It is also shown that the initial singularity appears even in other scenarios~\cite{Borde:1993xh,Borde:1996pt}. The incompleteness of the inflationary Universe strongly motivated us to explore non-singular scenarios in the very early Universe. The bouncing Universe (see, e.g.,  \cite{Brandenberger:2016vhg} for a review) is an interesting candidate of a non-singular beginning of the Universe. We can also consider completely periodic Universe as a possible beginning. A cyclic Universe and an ekpyrotic Universe studied as an alternative to the inflation~\cite{Khoury:2001wf,Steinhardt:2001vw, Gould:2019xwr, Ijjas:2019pyf} belong to this category. Related to the cyclic Universe, a cosmological realization of a time crystal, called time crystal Universe \cite{Bains:2015gpv,Easson:2016klq,Easson:2018qgr} has been studied.

In this paper, we seek a cyclic Universe as a past completion of an inflationary Universe.
Notice that, to construct a bouncing solution, we need to violate the null energy condition. Since a scalar field minimally coupled with gravity 
in a flat Universe satisfies the null energy condition, we need to consider more general situations. It is Horndeski theory~\cite{Horndeski:1974wa,Deffayet:2011gz,Kobayashi:2011nu}(see \cite{Kobayashi:2019hrl} for a recent review) that is the general class of scalar-tensor theory whose Euler-Lagrange equations of motion contain up to second order derivatives. Actually, bouncing cosmology in the Horndeski theory has been intensively investigated so far~\cite{Cai:2012va,Koehn:2013upa,Battarra:2014tga,Qiu:2015nha,Ijjas:2016tpn}. These analyses show the presence of gradient instability~\cite{Kawai:1998ab,Kawai:1999pw} and finally no-go theorem for stable bouncing cosmology in Horndeski theory was found for the spatially flat Universe~\cite{Kobayashi:2016xpl}. However, it was shown that no-go theorem does not hold when the spatial curvature is included~\cite{Akama:2018cqv}.
Therefore, we should consider Horndeski theory in the presence of a spatial curvature in order to have a bouncing Universe \cite{Matsui:2019ygj}.

Recently, it was found that a subclass of Horndeski theory possesses a dual expression described by a 2-form gauge field \cite{Yoshida:2019dxu}, as in the case of a free massless scalar field. In the case of the free massless field, the 2-form dual theory provides a topology changing tunneling process from a flat Euclidean space $\mathrm{R}^3$ to a direct sum space $S^3 \oplus \mathrm{R}^3$ which is mediated by the Giddings-Strominger instanton \cite{Giddings:1987cg}. The nucleated closed Universe $S^3$ is called baby Universe. It is known that the baby Universe contracts after nucleation.
 Hence, it does not represent the realistic expanding Universe. One possibility to circumvent this conclusion  would be to introduce an additional field which provides a vacuum energy \cite{Rubakov:1988wx}. As an alternative possibility, in this paper, we regard the contracting baby Universe as a contracting phase of a bouncing/cyclic Universe in Horndeski theory.
Since a spatial curvature is naturally introduced in this set up, there is a chance to realize a bouncing Universe.
 Indeed, we find a time crystal Universe as a baby Universe nucleated from a flat space, which is mediated by an extension of Giddings-Strominger instanton in a 2-form theory dual to the Horndeski theory.
 Interestingly, once a cosmological constant is introduced, it turns out that de Sitter Universe can be created from
  time crystal. Thus, we have a completion of inflationary scenario, namely, the past boundary of an inflationary Universe is   time crystal.

The paper is organized as follows. In the next section, we review the 2-form theory dual to a subclass of the shift symmetric Horndeski theory~\cite{Yoshida:2019dxu}. Then in the section \ref{sec3}, we construct a time crystal Universe in the 2-form dual of Horndeski theory. In the section \ref{sec4}, we solve Euclidean equation of motion in 2-form theory without a cosmological constant and construct an extension of Giddings-Strominger instanton. This instanton describes a nucleation of a time crystal Universe from a flat space. Then, in the section \ref{sec4B}, we extend this analysis in the presence of a cosmological constant and find an instanton solution which mediates a tunneling process from a time crystal Universe to an inflationary Universe. Thus, a past completion of an inflationary scenario is obtained. The final section is devoted to summary and discussion. There we comment on the stability issue.

\section{2-form dual of Horndeski theory}
In this section, we review a 2-form gauge theory dual to a subclass of Horndeski theory~\cite{Yoshida:2019dxu}. We refer the reader to the original paper \cite{Yoshida:2019dxu} for a detailed derivation. 

 We focus on a shift symmetric scalar field coupling to gravity through the Einstein tensor $G^{\mu\nu}$ as,
\begin{align}
S^{\phi} = \int d^4 x \sqrt{-g} \left[ \frac{M_{\mathrm{pl}}^2}{2} (R - 2 \Lambda) - \frac{\alpha}{2}g^{\mu\nu}\partial_{\mu}\phi \partial_{\nu} \phi - \frac{\beta}{2} G^{\mu\nu}\partial_{\mu}\phi \partial_{\nu}\phi \right], \label{Sphi}
\end{align}
where $\alpha$ and $\beta$ are constants. We assume $\alpha > 0$ to ensure that the kinetic term of the scalar field $\phi$ has a correct sign at the low energy. Though the Einstein tensor includes the second derivatives of the metric, equations of motion include up to the second derivatives. Hence, this interaction does not have the Ostrogradsky's ghost. Actually, this is a sub-class of Horndeski theory with $G_{2} = \alpha/2,$ $G_{4} = M_{\mathrm{pl}}^2/2 + \beta X$ and $G_{3} = G_{5} = 0$, where $X$ stands for $g^{\mu\nu}\partial_{\mu}\phi \partial_{\nu}\phi$. This class of theory was, for example, studied to construct hairy black hole solutions \cite{Babichev:2013cya}.  For convenience, we introduce the effective metric ${\cal G}^{\mu\nu}$ by
\begin{align}
 {\cal G}^{\mu\nu} = \alpha g^{\mu\nu} + \beta G^{\mu\nu}.
\end{align}
Then, the action can be written as
\begin{align}
S^{\phi} = \int d^4 x \sqrt{-g}\left[
\frac{M_{\mathrm{pl}}^2}{2} (R- 2 \Lambda)  - \frac{1}{2} {\cal G}^{\mu\nu} \partial_{\mu} \phi \partial_{\nu} \phi  \right] .
\end{align}

As is well known, a free massless scalar field is equivalent to a free 2-form gauge field through the duality, $d \phi = * d B $, where $B$ is a 2-form field and $*$ represents the Hodge dual.
In Ref. \cite{Yoshida:2019dxu}, it was shown that the similar duality holds even if the derivative coupling through the Einstein tensor \eqref{Sphi} is included. 
The resultant action is given by
\begin{align}
 S = \int d^4 x \sqrt{-g} \left[ \frac{M_{\mathrm{pl}}^2}{2}(R - 2 \Lambda) - \frac{1}{12} {\cal G}^{\mu\nu\rho}{}_{\alpha\beta\gamma} H_{\mu\nu\rho} H^{\alpha\beta\gamma} \right],\label{SH}
\end{align}
where $H = d B$ is the field strength of the 2-form gauge field $B_{\mu\nu}$. The components of $H$ are given by
\begin{align}
 H_{\mu\nu\rho} = 3 \partial_{[\mu} B_{\nu\rho]} = \partial_{\mu} B_{\nu\rho} + \partial_{\nu} B_{\rho\mu} + \partial_{\rho} B_{\mu\nu}.
\end{align}
The tensor ${\cal G}^{\mu\nu\rho}{}_{\alpha\beta\gamma}$ is defined by
\begin{align}
{\cal G}^{\mu\nu\rho}{}_{\alpha\beta\gamma} := \frac{{\cal G}^{\mu}{}_{\alpha} {\cal G}^{\nu}{}_{\beta} {\cal G}^{\rho}{}_{\gamma}}{\det {\cal G}^{\cdot}{}_{\cdot}},\label{calG}
\end{align}
where $\det{{\cal G}^{\cdot}{}_{\cdot}}$ is the determinant of ${\cal G}^{\mu}{}_{\nu}$. The duality relation is no longer given by the simple Hodge dual. It now depends on the curvature of spacetime as
\begin{align}
 {\cal G}_{\mu}{}^{\nu} \partial_{\nu} \phi = \frac{1}{3!} \epsilon_{\mu}{}^{\nu\rho\sigma} H_{\nu\rho\sigma}.
\end{align} 

In the next section, we study the above action and obtain an extension of the Giddings-Strominger instanton, which is an instanton solution in free 2-form theory with gravity.

\section{Time Crystal Universe with 2-form charge} 
\label{sec3}
In this section, we construct a cosmological solution of the 2-form theory \eqref{SH}. Note that, thanks to the duality, following results can be obtained even if one starts with the scalar action \eqref{Sphi}. The essential difference will appear when we discuss quantum effects in the next section.  

We consider homogeneous and isotropic ansatz without specifying the spatial curvature,
\begin{align}
 g_{\mu\nu}dx^{\mu} dx^{\nu} &= l^2 \left[ - N(t)^2 dt^2 + a(t)^2 \Omega_{ij} dx^{i} dx^{j} \right],
\end{align}
where $\Omega$ is given by
\begin{align}
 \Omega_{ij}dx^{i} dx^{j} = d\chi^2 + f_{k}(\chi)^2 (d \theta^2 + \sin^2 \theta d\phi^2), \qquad f_{k}(\chi) = 
\begin{cases}
 \sin \chi &(k = + 1)\\
 \chi  &(k = 0) \\
 \sinh \chi &(k = -1)
\end{cases}.\label{Omega}
\end{align}
Here $k$ represents the sign of the  spatial curvature.
We have introduced a free parameter $l$ with a mass dimension $-1$ so that the coordinates $(t, \chi, \theta, \phi)$ and the dynamical variables $N(t)$ and $a(t)$ are dimensionless. We will also use the (dimensionless) proper time $\tau$, that is defined by $d \tau = N d t$.

We assume homogeneous and isotropic configuration of the field strength $H$; 
\begin{align}
 H = h \sqrt{\Omega} d\chi \wedge d\theta \wedge d\phi  \ ,   \label{solH} 
\end{align}
where $h$ represents the magnetic flux density of the 2-from gauge field.  
The symmetry allows $h$ to depend on time but it is actually forbidden from the Bianchi identity $d H = 0$. One can check that our ansatz \eqref{solH} satisfies the equation of motion as well:
\begin{align}
 \nabla_{\mu} \left( {\cal G}^{\mu\nu\rho}{}_{\alpha\beta\gamma} H^{\alpha\beta\gamma} \right) = 0.
\end{align}
Now, we fix the free parameter $l$ by 
\begin{align}
  l := \left( \frac{h^2}{ 6 \alpha M_{\mathrm{pl}}^2}\right)^{1/4}.
\label{defaGS}
\end{align}
Then we introduce a dimensionless ratio of $\beta$ to $l^2 \alpha$, say $\gamma$, by
\begin{align}
 \gamma := \frac{3}{\alpha}\frac{\beta}{l^2} = \sqrt{\frac{6}{\alpha}}\frac{ 3 \beta M_{\mathrm{pl}}}{|h|},
\end{align}
and a dimensionless cosmological constant $\lambda$ by
\begin{align}
 \lambda := l^2 \Lambda.
\end{align}
Now, we obtain the mini-superspace action
\begin{align}
 S &=  3 l^2 M_{\mathrm{pl}}^2 {\cal V}  \int dt N a^3 \left[
\frac{\partial_{\tau}^2 a}{a} +
\left( \frac{\partial_{\tau} a}{a} \right)^2 + \frac{k}{a^2} - \frac{\lambda}{3} - \frac{1}{a^6} \frac{1}{1 - \gamma^2 ( (\partial_{\tau} a/a)^2 + k/a^2)}
\right],
\end{align}
where ${\cal V}$ is the comoving volume of constant time slices, ${\cal V} := \int d^3 x   \sqrt{\Omega}$.
Taking the variation of the action with respect to $N$, 
we obtain the modified Friedmann equation, 
\begin{align}
(\partial_{\tau}a)^2 + k - \frac{\lambda}{3} a^2 + \frac{- a^2 + \gamma (3 (\partial_{\tau}a)^2 + k)}{(a^3 - a \gamma ((\partial_{\tau}a)^2 + k))^2} = 0.\label{Friedmann}
\end{align}
By solving it for $(\partial_{\tau} a)^2$, we derive the convenient form
\begin{align}
 (\partial_{\tau}a)^2 = - V(a;\gamma,k,\lambda), \label{adot=V}
\end{align} 
where the effective potentials $V$ has 3 branches corresponding to 3 roots of the cubic equation;
\begin{align}
 V_{n}(a;\gamma,k,\lambda)
=
k
- \frac{6 + \lambda \gamma}{9 \gamma} a^2
+ \mathrm{e}^{- i \frac{2 \pi}{3}(n-1)} \frac{\Xi^{1/3}}{9 a \gamma^2}
+ \mathrm{e}^{ i \frac{2 \pi}{3}(n-1)}  \frac{- 81 \gamma + a^6 ( -3 + \gamma \lambda)^2}{9 a \Xi^{1/3}}, \qquad (n = 1,2,3) \label{defV}
\end{align}
with
\begin{align}
\Xi(a;\gamma,k,\lambda)/\gamma^3 =&
- 729 \gamma^2 k a - a^9 \left( -3 + \gamma \lambda
   \right)^3  + \frac{243}{2} \gamma a^3 \left(3 + \gamma  \lambda \right) \notag\\
 & +\frac{27}{2 \gamma} \Biggl[  \gamma^3 \Bigl( 2916 \gamma^2+2916 \gamma^3 k^2 a^2 + 8 \gamma k a^{10} \left(- 3 + \gamma \lambda \right)^3 - 972 \gamma^2 k a^4 \left(3 + \gamma \lambda \right)\notag\\
&\qquad \qquad    -8 a^{12} \left(- 3 \gamma \lambda \right)^3 -27 \gamma a^6 \left(\gamma ^2 \lambda^2-42 \gamma  \lambda +9\right) \Bigr) \Biggr]^{1/2}.
\end{align}
Here we define a branch cut of the power function as $z^{n} := |z|^n \mathrm{e}^{i n \mathrm{Arg} z} , \mathrm{Arg} z \in (-\pi , \pi]$.

\begin{figure}[thbp]
\begin{minipage}[cbt]{0.495\hsize}
\begin{center}
 \includegraphics[width=\hsize]{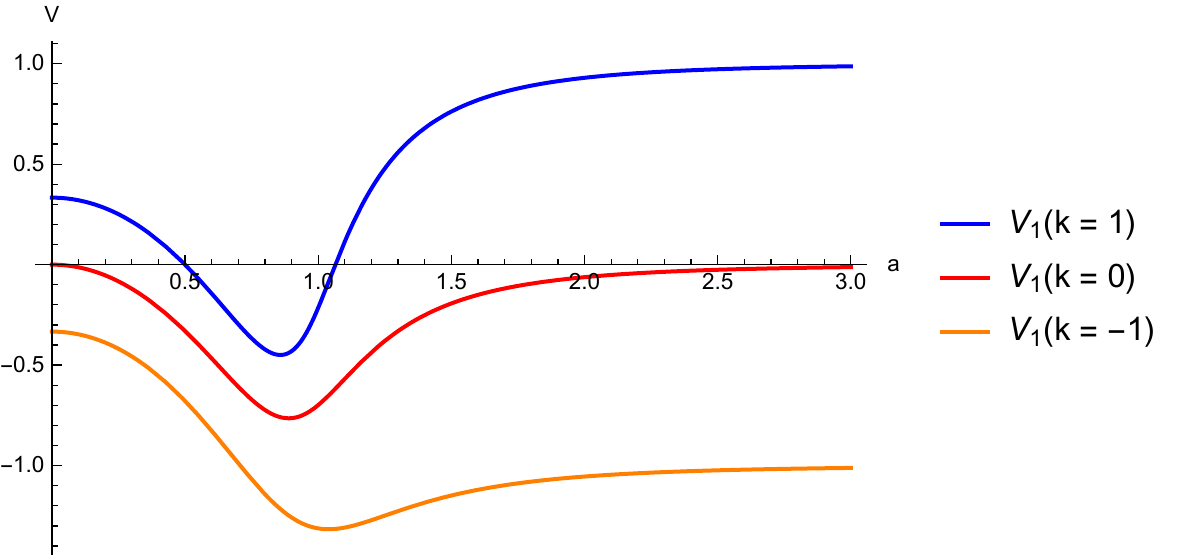}
\end{center}
\end{minipage}
\begin{minipage}[cbt]{0.495\hsize}
\begin{center}
 \includegraphics[width=\hsize]{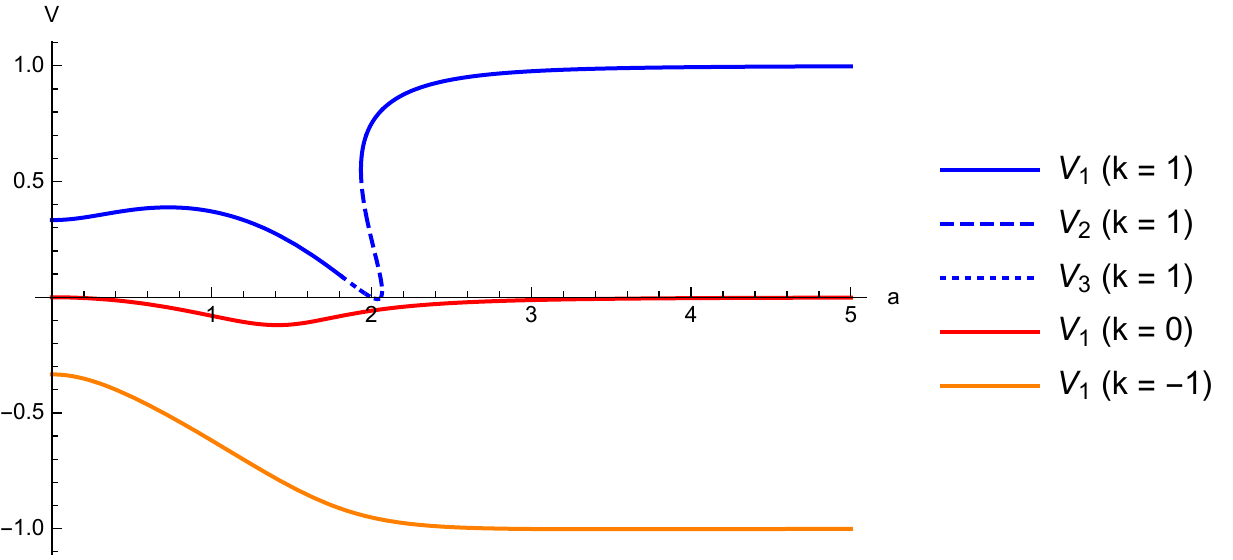}
\end{center}
\end{minipage}
 \caption{Potential for positive $\gamma$ (left: $ \gamma = \frac{1}{4}$, right: $\gamma = 4$) with $\lambda = 0$: The blue plot in the left figure provides an oscillating solution between $a_{\mathrm{min}} \sim 0.5$ and $a_{\mathrm{max}} \sim 1.08$. $V_{n}$ which are not written in the plot are complex value.}
\label{figVL1}
\end{figure}

\begin{figure}[htbp]
\begin{minipage}[cbt]{0.495\hsize}
\begin{center}
 \includegraphics[width=\hsize]{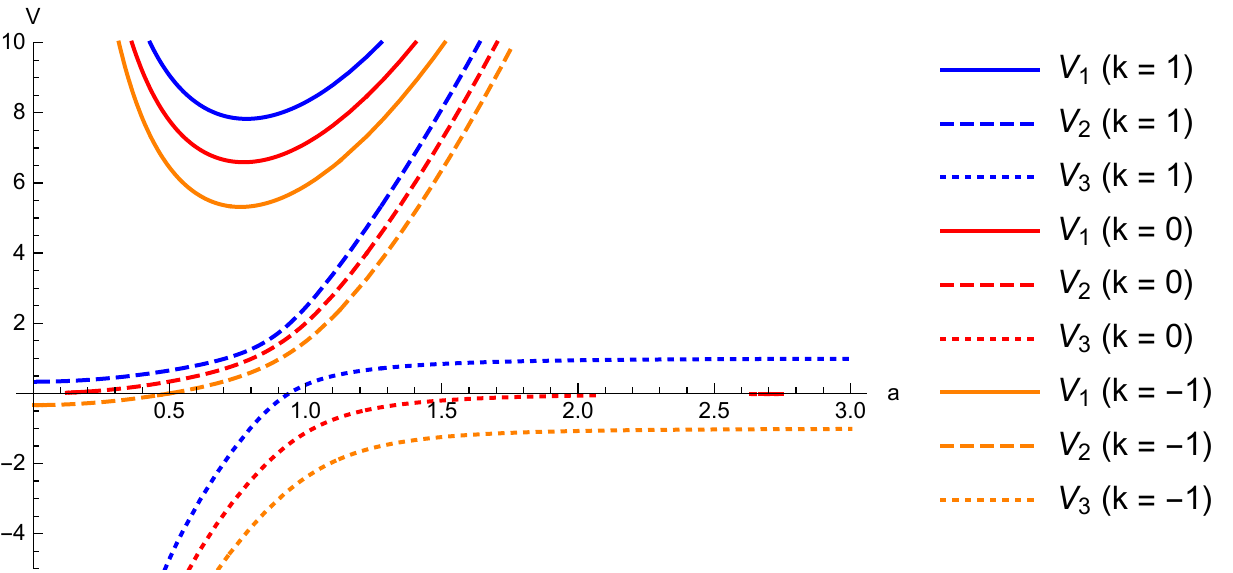}
\end{center}
\end{minipage}
\begin{minipage}[cbt]{0.495\hsize}
\begin{center}
 \includegraphics[width=\hsize]{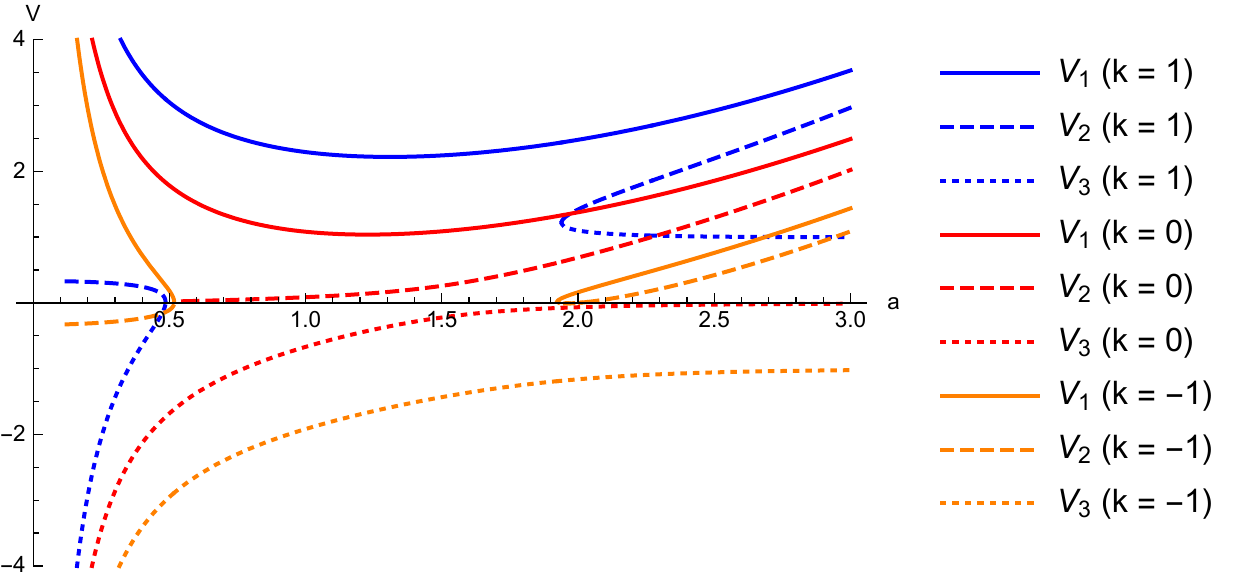}
\end{center}
\end{minipage}
 \caption{Potential for negative $\gamma$ (left: $ \gamma = -\frac{1}{4}$, right: $\gamma = -4$) with $\lambda = 0$: No potential has an oscillating solution.}
\label{figVL3}
\end{figure}
The behavior of the solution can be understood by the analogy of the dynamics of a point particle: Eq. \eqref{adot=V} is nothing but the energy conservation law of a point particle with the potential $V_{n}$ and a total energy $E = 0$. The typical potentials $V_{n}$ for $ \lambda = 0$ are plotted in fig.~\ref{figVL1} (positive $\gamma$) and fig.~\ref{figVL3} (negative $\gamma$). 
The most interesting solution is that of $k = 1$ with small positive $\gamma$ (the blue plot in the left of fig.~\ref{figVL1}), where the potential has 2 turning points $a = a_{\mathrm{min}}$ and $a_{\mathrm{max}}$, which are $0.5$ and $1.08$ in fig.~\ref{figVL1} respectively. The scale factor will oscillate between $a_{\mathrm{min}}$ and $a_{\mathrm{max}}$. Thus the solution is expected to express a time crystal Universe, where scale factor is exactly periodic. We will investigate this solution analytically in the next subsection. Note that any potential other than $k =1$ with small positive $\gamma$ has at most one terming point and does not have a time crystal solution.

\begin{figure}[htbp]
\begin{center}
 \includegraphics[width=10cm]{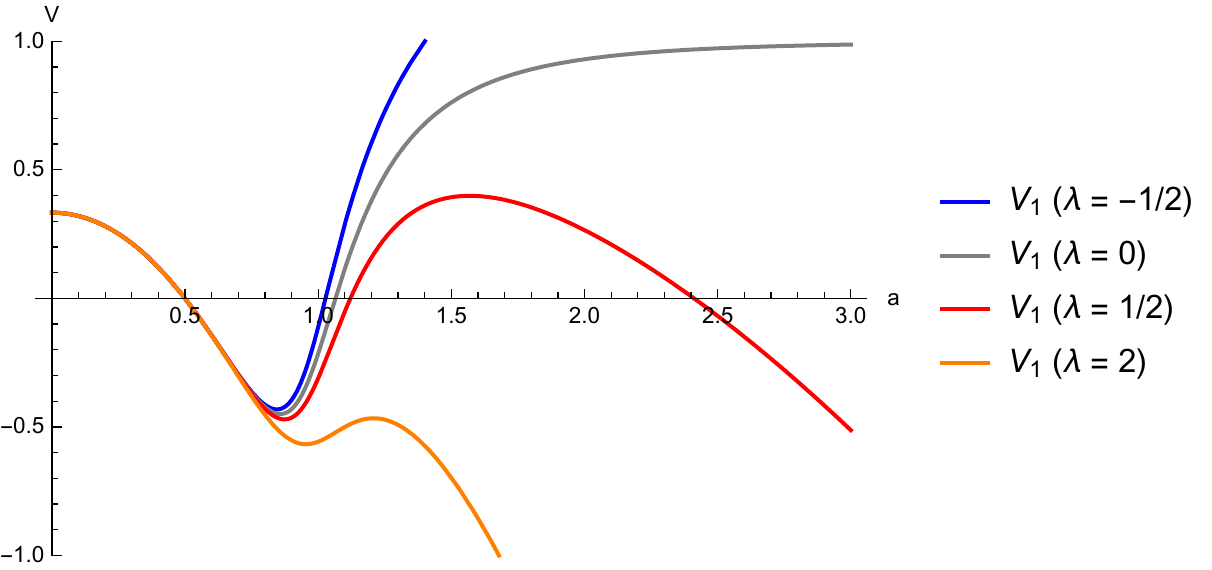}
 \caption{Potential with cosmological constant: $\gamma = \frac{1}{4}$: A small cosmological constant (red plot) provides a new branch of solution $a > a_{\Lambda} \sim 2.4$ in addition to the oscillating solution $a \in (a_{\mathrm{min}} \sim 0.5, a_{\mathrm{max}} \sim 1.12)$.}
\label{figVL5}
\end{center}
\end{figure}
Once the cosmological constant $\lambda$ is turned on, the potential changes as in fig.~\ref{figVL5}.
For a negative $\lambda$ (blue), there is no essential difference from the $\lambda = 0$ case.
For a large positive  $\lambda$ (orange), the time crystal solution no longer exists and Universe enters the $\lambda$ dominant eta. For a small positive  $\lambda$ (red), the potential has three zero points, $a = a_{\mathrm{min}}, a_{\mathrm{max}}$ and $a_{\Lambda}$, 
 $a \sim 0.5, 1.12$ and $2.4$ in fig.~\ref{figVL5}. The time crystal Universe, which oscillates between $a_{\mathrm{min}}$ and $a_{\mathrm{max}}$, still exist. In addition, there is another solution in $a > a_{\Lambda}$. Since the cosmological constant dominates over any other term as $a$ grows, this solution asymptotically approaches a de Sitter spacetime.
In the section \ref{sec4B}, we will see that the time crystal Universe can make a transition quantum mechanically into the inflationary Universe. 

\subsection{analytic solution}
We would like to construct the analytic solution for the case $k = 1$, $\lambda = 0$, and a small positive $\gamma$.
It is useful to find the analytic formulas for $a_{\mathrm{min}}$ and $a_{\mathrm{max}}$ by solving 
Eq.\eqref{adot=V} with setting $\partial_{\tau}a = 0$,
\begin{align}
a_{\mathrm{min}} &:= \sqrt{\gamma}, \qquad a_{\mathrm{max}} := \sqrt{\frac{\gamma + \sqrt{4+\gamma^2}}{2}},\label{ahatminmax}
\end{align}
which corresponds to the minimum and maximal size of oscillating Universe. 
Since a time coordinate has not fixed yet, we can choose it by fixing a functional form of $a(t)$ freely.
Our choice of the time coordinate $t$ is as follows:
\begin{align}
 a(t) := \frac{(a_{\mathrm{min}} + a_{\mathrm{max}}) - (a_{\mathrm{min}} - a_{\mathrm{max}})\cos 2t}{2}. \label{sola}
\end{align}
Then the lapse function associated with this time coordinate is given by plugging \eqref{sola} into \eqref{adot=V}:
\begin{align}
 N(t) = (a_{\mathrm{max}} - a_{\mathrm{min}})\sqrt{\frac{\sin^2 2t}{-V_{1}(a(t))}}. \label{solN}
\end{align}
Note that the analytic expression of $V_{1}$ is given by Eq. \eqref{defV}. The equations \eqref{sola} and \eqref{solN} provide an analytic solution, which is plotted in fig.~\ref{aNplot}:
\begin{figure}[htbp]
\begin{center}
 \includegraphics[width=10cm]{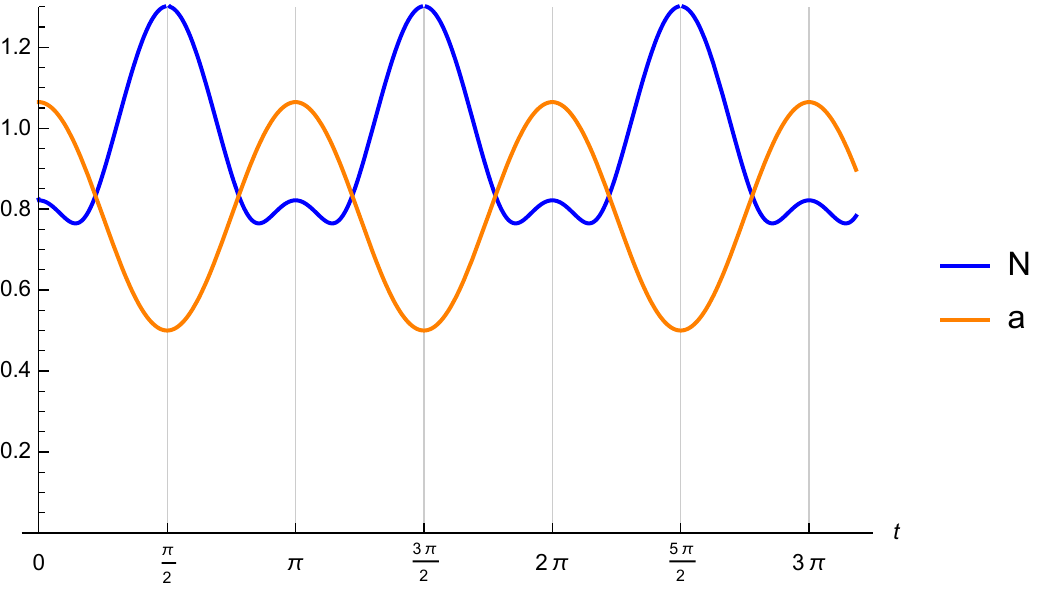}
 \caption{Plot of the lapse function $N$ and the scale factor $a$}
\label{aNplot}
\end{center}
\end{figure}

Note that if the choice of time coordinate \eqref{sola} were ill defined at some point, 
a singular behavior appears in $N(t)$, which tell us a presence of coordinate singularity. 
For example, if one choose a time coordinate $t$ by $a = t$, a positive $N^2$ can be obtained only for $a_{\mathrm{min}} < t < a_{\mathrm{max}}$ and $t = a_{\mathrm{min}}$ and  $a_{\mathrm{max}}$ correspond to coordinate singularities. The numerical plot in fig.~\ref{aNplot} shows that our choice of a time coordinate \eqref{sola} has no coordinate singularity.

\begin{figure}[htbp]
\begin{minipage}{0.49\hsize}
\begin{center}
 \includegraphics[width=\hsize]{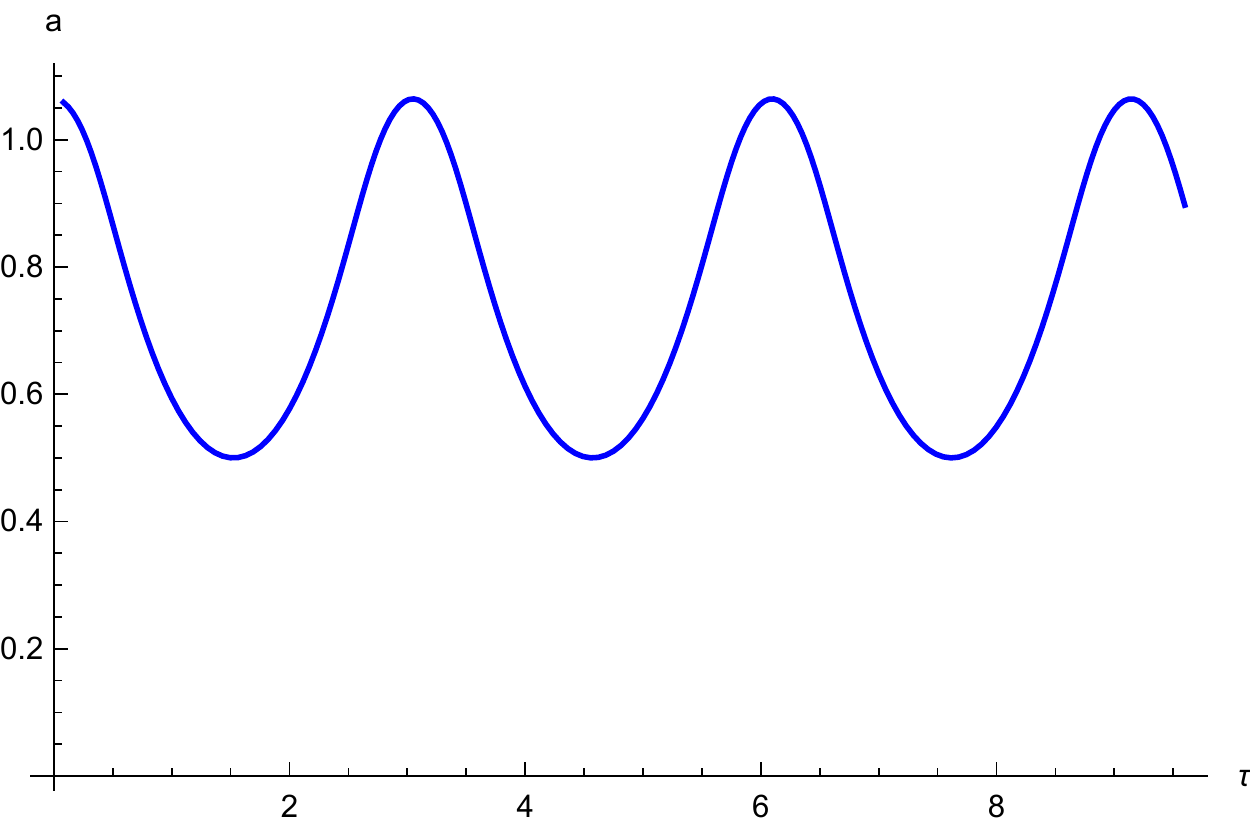}
 \caption{The scale factor $a$ as a function of the proper time $\tau$: scale factor $a$ is a periodic function of $\tau$.}
\label{atauplot}
\end{center}
\end{minipage}
\begin{minipage}{0.49\hsize}
\begin{center}
 \includegraphics[width=\hsize]{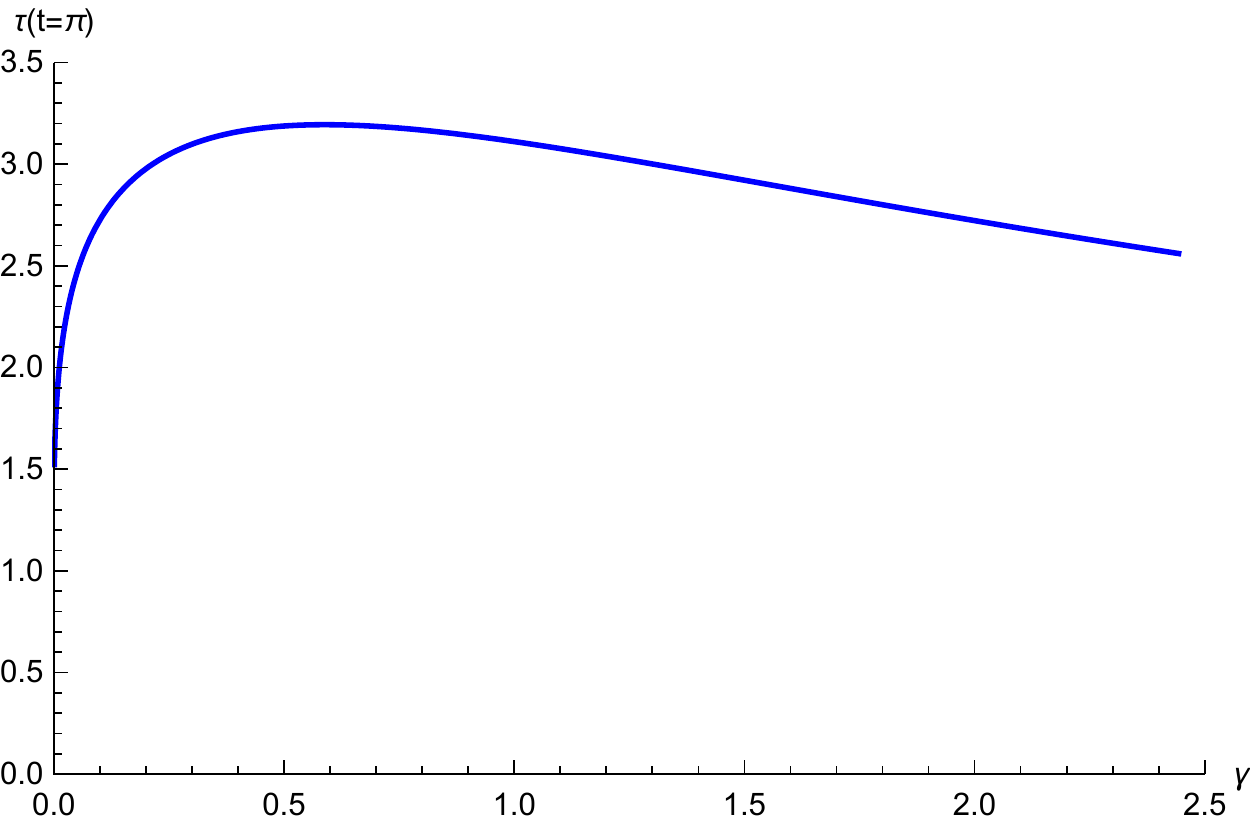}
 \caption{$\gamma$ dependence of the period of oscillation measured by the proper time $\tau$}
\label{fig:taugamma}
\end{center}
\end{minipage}
\end{figure}
\begin{figure}[b]
\begin{center}
 \includegraphics[width=0.55\hsize]{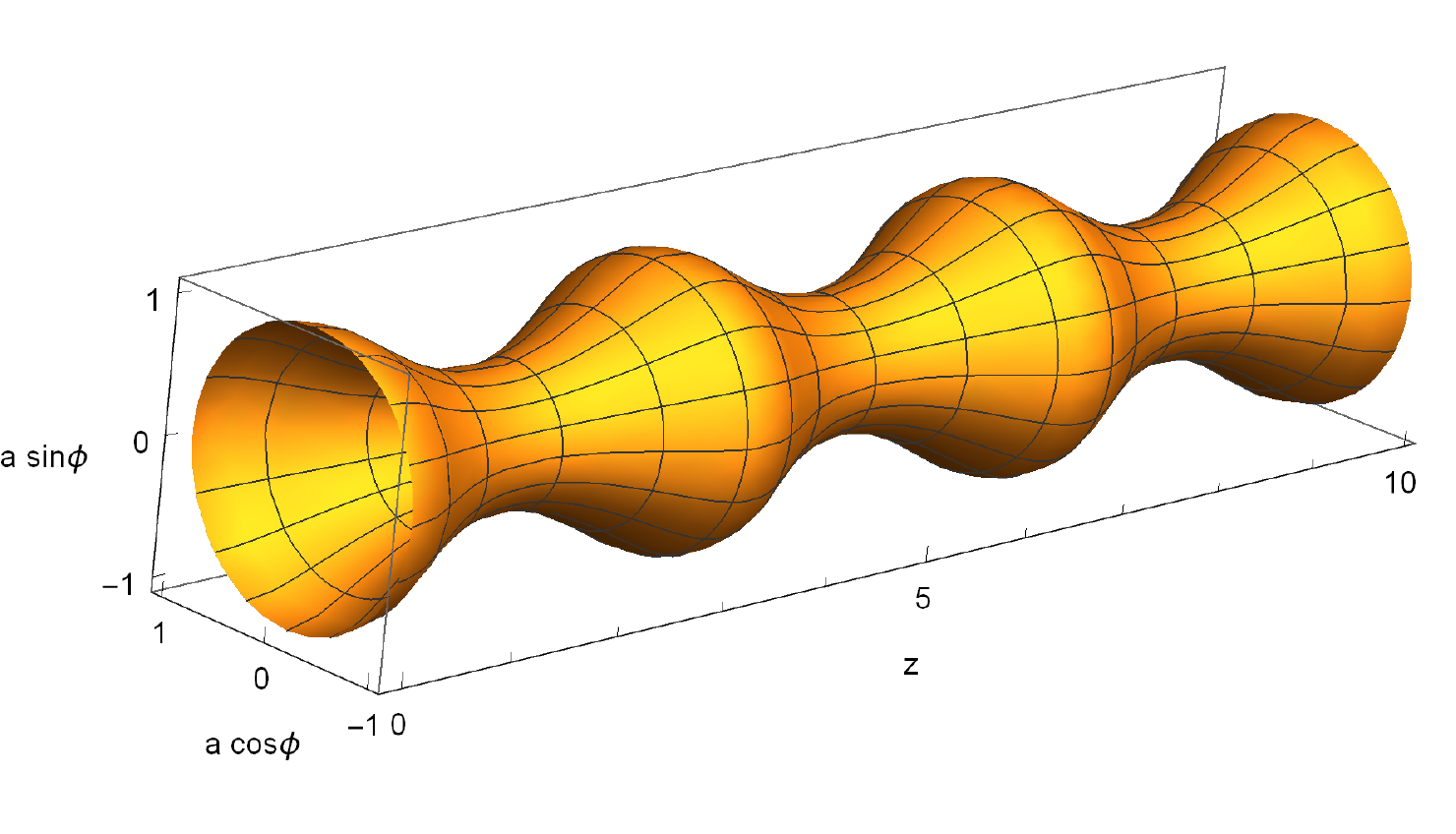}
 \caption{2-dimensional time crystal Universe (setting $\chi = \theta = \pi/2$) embedded in 3-dimensional Minkowski spacetime with a time coordinate $z$.}
\label{figTCU}
\end{center}
\end{figure}

Since the expression of our solution in fig.~\ref{aNplot} highly depends on our choice of a time coordinate, it is useful to express our result by the proper time.  The proper time $\tau$  can be obtain by integrating $N$,
\begin{align}
 \tau(t) = \int^{t}_{0} d t N.
\end{align}
The scale factor can be written as a function of this proper time, which is plotted in fig.~\ref{atauplot}. 
Here the period of scale factor in terms of the proper time is $\tau(t = \pi) \sim 3.05$ when $\gamma = \frac{1}{4}$. The proper time $\tau(\pi)$ depends on the value of $\gamma$ as shown in fig.~\ref{fig:taugamma} .

As another coordinate independent expression of our result, it is useful to embed the 4-dimensional time crystal Universe to the 5-dimensional Minkowski spacetime,
\begin{align}
l^2 \left[ - dz^2 + d a^2 + a^2 \Omega_{ij}dx^{i}dx^{j} \right],
\end{align}
by the embedding function $z = z(t)$ and $a = a(t)$.  Here, $a(t)$ is the scale factor of the time crystal Universe while the function $z = z(t)$ can be determined by imposing that the pull back of 5-dimensional Minkowski metric coincides with our time crystal Universe, that is, 
\begin{align} 
l^2 \left[ -( (\partial_{t}z(t))^2 - (\partial_{t}a)^2) dt^2 + a(t)^2 \Omega_{ij}dx^{i}dx^{j} \right] = l^2 \left[
- N^2 dt^2 + a^2(t) \Omega_{ij}dx^{i}dx^{j}
\right].
\end{align}
Then it can be integrated as
\begin{align}
 z(t) = \int^{t}_{0} dt \sqrt{N^2 + (\partial_{t}a)^2},
\end{align}
where the initial condition at $t = 0$ is set to the origin of $z$.
Actually, we can visualize this embedding as a 2-dimensional surface in the 3-dimensional Minkowski spacetime by considering a subspace $\chi = \theta = \frac{\pi}{2}$, where $\chi$ and $\theta$ are the polar coordinates given by Eq.\eqref{Omega} with $k = +1$. 
The embedding can be seen in fig.~\ref{figTCU}.  Embedding the solution  into the flat space will help us to understand the analytic continuation of an Euclidean solution to a Lorentzian solution as we will see in the following sections.

\section{Nucleation of time crystal Universe from flat space}
\label{sec4}
\label{sec4A}
Since the duality is an on-shell relation, quantum effects such as tunneling effect are expected to show differences. The Euclidean solution with the free 2-form field is known as a Giddings-Strominger instanton \cite{Giddings:1987cg}, which  provides a topology changing tunneling process from a flat space $\mathrm{R}^3$ to a sum of a flat space $\mathrm{R}^3$ and a closed FLRW Universe $S^3$ called a baby Universe. In following, we will investigate how the Giddings-Strominger instanton is modified in the 2-form dual of Horndeski theory.    

To construct an instanton solution, we first need to perform the Wick rotation, $t \rightarrow - i t_{E}$, to obtain an Euclidean action:
\begin{align}
 S_{E} = - i S|_{t \rightarrow - i t_{E}} 
= \int dt_{E} d^3x \sqrt{g_{E}} \left[ -\frac{M_{\mathrm{pl}}^2}{2}(R_{E} - 2 \Lambda) + \frac{1}{12} {\cal G}_{E}{}^{\mu\nu\rho,\alpha\beta\gamma} H_{\mu\nu\rho} H_{\alpha\beta\gamma}\right]  \label{SEH}.
\end{align}
Here $g^{E}_{\mu\nu}$ is the Euclidean metric, $R^{E}$ and $G^{E}_{\mu\nu}$ are the Ricci scalar and the Einstein tensor with respect to $g^E_{\mu\nu}$. ${\cal G}_{E}$ is the Euclidean version of \eqref{calG}, which is defined by  
 \begin{align}
 {\cal G}_{E}{}^{\mu\nu\rho,\alpha\beta\gamma} := \frac{{\cal G}_{E}{}^{\mu \alpha} {\cal G}_{E}{}^{\nu \beta} {\cal G}_{E}{}^{\rho \gamma}}{\det {\cal G}_E{}_{\cdot}{}^{\cdot}},
\qquad 
  {\cal G}_{E}{}^{\mu \nu} = \alpha g_{E}^{\mu \nu} + \beta G_{E}{}^{\mu\nu} .
\end{align}

Let us consider Euclidean version of the FLRW ansatz, 
\begin{align}
 g^{E}_{\mu\nu}dx^{\mu} dx^{\nu} &= l^2 \left[ N(t_{E})^2 d t_{E}^2 + a(t_{E})^2 \Omega_{ij} dx^{i} dx^{j} \right], \label{E-FLRW} 
\end{align}
with a constant magnetic flux of 2-form field,
\begin{align}
 H = h \sqrt{\Omega}\ d\chi \wedge d\theta \wedge d\phi. \label{H=}
\end{align}
As similar to the Lorentzian case, $H$ satisfies the field equation
\begin{align}
 \nabla_{\mu} \left( {\cal G}_{E}{}^{\mu\nu\rho}{}_{\alpha\beta\gamma} H^{\alpha\beta\gamma} \right) = 0,
\end{align}
 as well as the Bianchi identity $d H = 0$.
Here we again set a length scale $l$ by Eq.\eqref{defaGS}.
Above field equation was derived by taking variation of the Euclidean action
with respect to  $B_{\mu\nu}$ as
\begin{align}
0 = \delta_{B} S_{E} = \int dt_{E} d^3 x \sqrt{g_{E}} \left[ - \frac{1}{2} \nabla_{\mu} \left( {\cal G}_{E}{}^{\mu\nu\rho}{}_{\alpha\beta\gamma} H^{\alpha\beta\gamma} \right) \delta B_{\nu\rho}
 + \frac{1}{2} \nabla_{\mu} \left( {\cal G}_{E}{}^{\mu\nu\rho}{}_{\alpha\beta\gamma} H^{\alpha\beta\gamma} \delta B_{\nu\rho}\right)
 \right].
\end{align}
Note that the second term automatically vanishes under our ansatz \eqref{E-FLRW} and \eqref{H=}. That ensures that we do not need to introduce an additional boundary term with respect to the 2-form field even for a boundary condition $\delta B_{\mu\nu} \neq 0$.

Now the  Euclidean action for the mini-superspace can be obtained as
\begin{align}
 S_{E} = 3 l^2 M_{\mathrm{pl}}^2 {\cal V} \int dt_{E} N a^3 \left[
\frac{\partial_{\tau_{E}}^2 a}{a} +
\left( \frac{\partial_{\tau_{E}} a}{a} \right)^2 - \frac{k}{a^2} + \frac{\lambda}{3} + \frac{1}{a^6} \frac{1}{1 + \gamma^2 ( (\partial_{\tau} a/a)^2 - k/a^2)}
\right].
\end{align}
Note that our new interaction term does not include the second derivative of the scale factor. That means we only need the Gibbons-Hawking term of the Einstein-Hilbert term for gravitational variation in mini-super space.
By taking the  variation of the  Euclidean action $S_{E}$ with respect to $N$, we can deduce the modified Euclidean Friedmann equation as
\begin{align}
 0 &=  - (\partial_{\tau_{E}} a)^2 + k - \frac{\lambda}{3} a^2 + \frac{- a^2 + \gamma ( - 3 (\partial_{\tau_{E}}a)^2  + k )}{\left(a^3 - \gamma a(- (\partial_{\tau_{E}}a)^2 + k)\right)^2} .\label{FriedmannEH}
\end{align} 
From this expression, we see that the Euclidean Friedmann equation can be obtained from Lorentzian Friedmann equation simply by replacing $(\partial_{\tau} a)^2 \rightarrow - (\partial_{\tau_{E}} a)^2$. Thus by defining the potential for the scale factor by 
\begin{align}
 (\partial_{\tau_{E}} a)^2 = - V^{E}_{n} (a; \gamma, \lambda, k), \qquad n = 1, 2, 3
\end{align}
we find that the Euclidean potential is nothing but the negative signed Lorentzian potential, 
\begin{align}
 V_{n}^{E} (a; \gamma, \lambda, k) = - V_{n}(a; \gamma, \lambda, k).
\end{align}
We note that this is because the 2-from field contribute to the Friedmann equation through a gradient energy, not a kinetic energy. In the case of scalar theory, the contribution of the scalar field to the Friedmann equation is from the kinetic term and the last term in Eq.\eqref{FriedmannEH} receives an additional minus sign.

\begin{figure}[htbp]
\begin{center}
 \includegraphics[width=12cm]{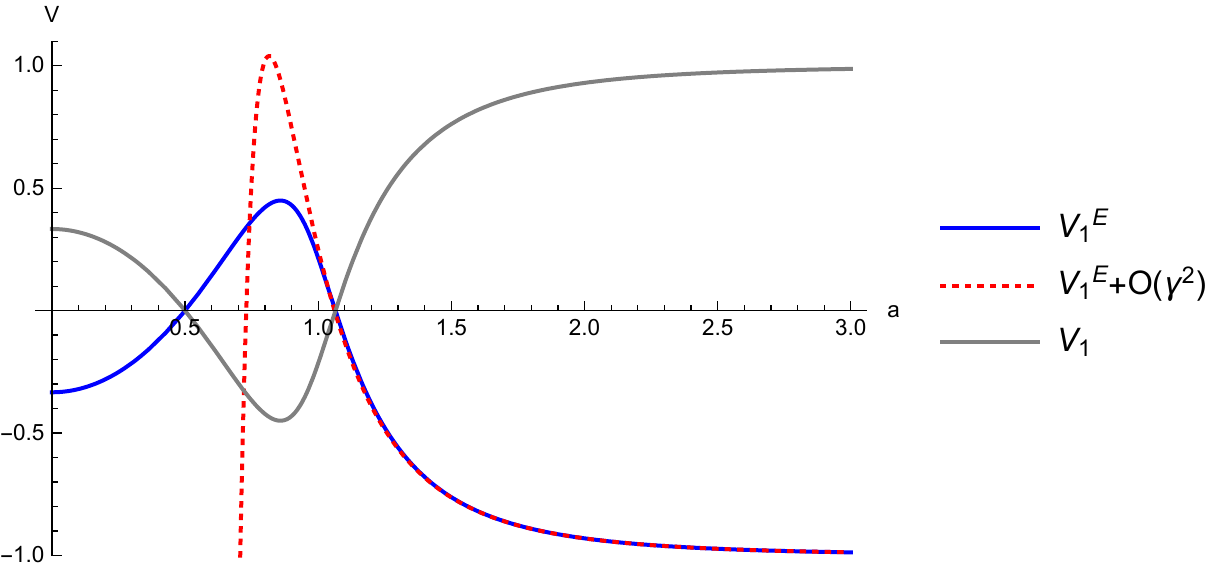}
 \caption{Euclidean potential $V^{E}$ in 2-form theory without cosmological constant ($\gamma = \frac{1}{4}$, $k = 1$, $\lambda = 0$):
The gray plot is the Lorentzian potential for time crystal $a \in (a_{\mathrm{min}} \sim 0.5, a_{\mathrm{max}} \sim 1.08)$. The red, dashed plot is a Taylor expansion of $V_1^{E}$ given by \eqref{VEOgamma2}.
}
\label{VE2}
\end{center}
\end{figure}
The Euclidean potential for $\lambda = 0$ case is plotted in fig.~\ref{VE2}. There are 2 possible solutions, $a \in (0, a_{\mathrm{min}})$ and $a \in (a_{\mathrm{max}}, \infty)$, where $a_{\mathrm{min}}$ and $a_{\mathrm{max}}$ are minimum and maximum size of the Lorentzian time crystal Universe defined by \eqref{ahatminmax}. 
We are interested in the outer solution $a \in (a_{\mathrm{max}}, \infty)$ because it corresponds to the Giddings-Strominger instanton.

As seen from the fig.~\ref{VE2}, in the $a > a_{\mathrm{max}}$ region, the potential can be approximately expressed by the Taylor expansion with respect to $\gamma$,
\begin{align}
 V^{E}_{1}(a;\gamma, \lambda = 0, k = 1 ) = - 1 + \frac{1}{a^4} - \frac{1}{a^6} \left(
- 2 + \frac{1}{a^4} \right) \gamma + {\cal O}(\gamma^2).\label{VEOgamma2}
\end{align}
As in the Lorentzian case, we can choose a time coordinate by fixing the functional form of $a(t_{E})$.
Our choice of the time coordinate is the following
\begin{align}
 a(t_{E}) := a_{\mathrm{max}} \sqrt{\cosh t_{E}}.
\end{align}
Then the Friedmann equation provides the expression for $N$ as
\begin{align}
 N(t_{E})^2 = \frac{1}{4} \cosh t_{E} + \frac{1}{8} \left(\cosh t_{E} - \frac{2}{1 + \cosh t_{E}} + 2 \frac{1}{\cosh^2 t_{E}}\right) \gamma + {\cal O}(\gamma^2).
\end{align}
Thus we obtain the solution,
\begin{align}
 g^{E}_{\mu\nu}dx^{\mu} dx^{\nu}
= l^2 \cosh \left( t_E \right) \left[ \left(  \frac{1}{4} + \frac{\gamma}{8} \left(1 - \frac{2}{\cosh(t_{E})(1 + \cosh t_{E})} + \frac{2}{\cosh^3 t_{E}}\right) \right) d t_{E}^2 + \left(1 + \frac{1}{2} \gamma\right)\Omega_{ij}dx^{i}dx^{j}
\right] + {\cal O}(\gamma^2). \label{WH}
\end{align}
Note that the leading term ( $\gamma = 0$ ) is nothing but the Giddings-Strominger solution.
\begin{figure}[htbp]
\begin{minipage}{0.49\hsize}
\begin{center}
 \includegraphics[width=\hsize]{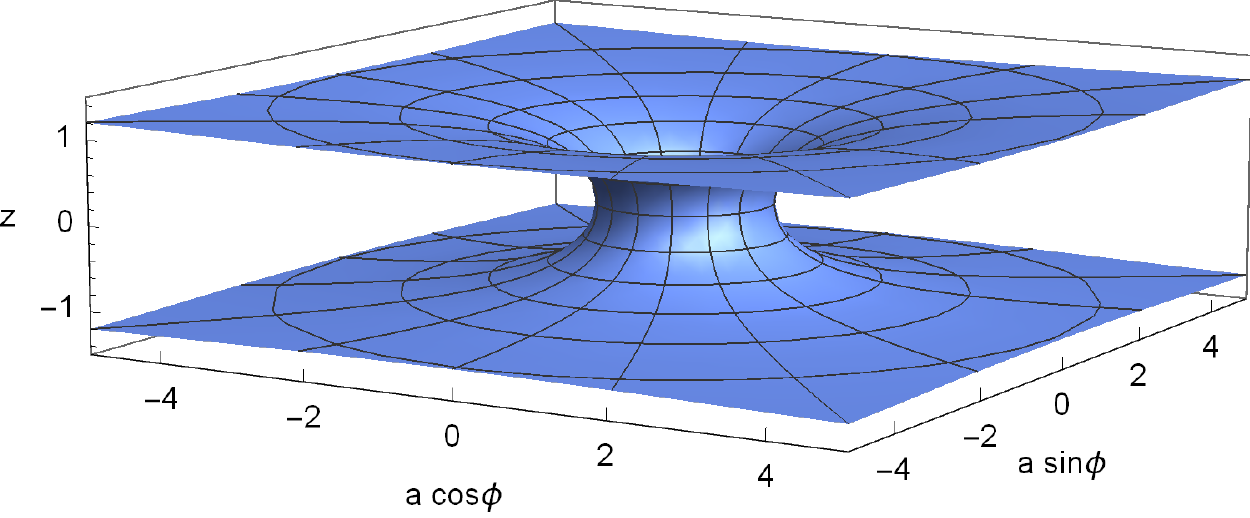}
 \caption{2-dimensional wormhole space embedded in 3-dimensional Euclid space: Each circles corresponds to $t_{E}$ constant curve. There are two asymptotic flat region (upper and lower planes).  }
\label{GSWH}
\end{center}
\end{minipage}
\begin{minipage}{0.49\hsize}
\begin{center}
 \includegraphics[width=\hsize]{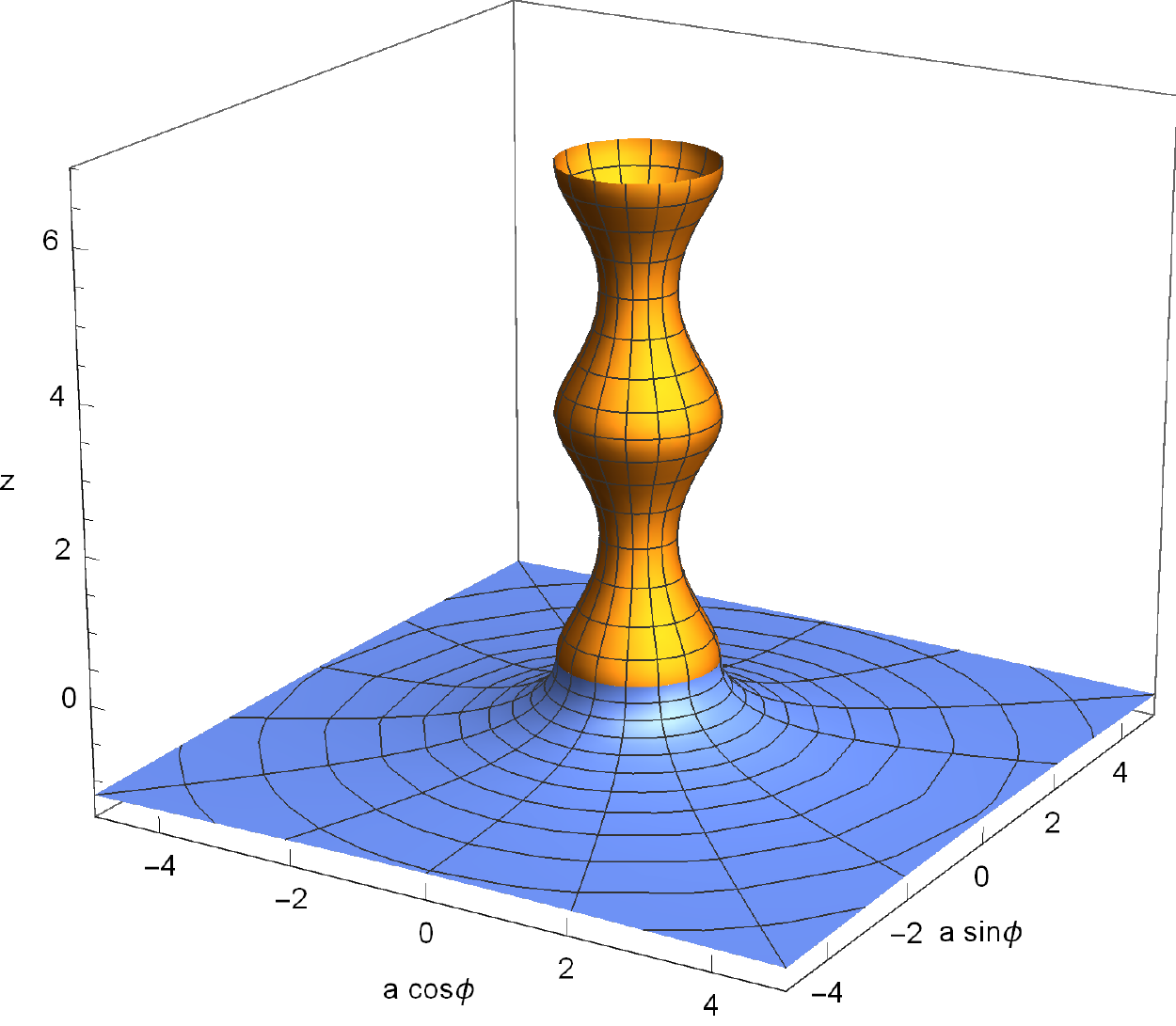}
 \caption{Analytic continuation of a Lorentzian time crystal Universe (orange) from a wormhole space (blue): Here $z>0$ is Minkowski space and $z<0$ is Euclid space. The two sides of asymptotic flat regions (e.g. $T := a \cos \phi  \rightarrow \pm \infty$) can also be connected the Lorentzian Minkowski spacetimes by the analytically continuation based on Cartesian coordinates $T \rightarrow i T$.}
\label{AC1}
\end{center}
\end{minipage}
\end{figure}
In the asymptotic region $t_{E} \rightarrow \pm \infty$, our metric reduces that of the flat space
\begin{align}
 g^{E}_{\mu\nu} dx^{\mu}dx^{\nu} &\rightarrow l^2 \left(1 + \frac{\gamma}{2} +{\cal O}(\gamma^2)\right) \left[
 \frac{1}{4} \mathrm{e}^{t_{E}} dt_{E}^2 + \mathrm{e}^{t_{E}} \Omega_{ij}dx^{i}dx^{j} 
\right] 
= l^2 \left[ d \tau_{E}^2 + \tau^2_{E} \Omega_{ij}dx^{i}dx^{j} \right], 
\end{align}
where the proper time $\tau_{E}$ is given as
\begin{align}
 \tau_{E} =  \left(1 + \frac{\gamma}{2} +{\cal O}(\gamma^2)\right)^{1/2} \mathrm{e}^{t_{E}/2}.
\end{align}
The metric \eqref{WH} represents a wormhole space which connects two distinct asymptotic flat spaces $t_{E} \rightarrow \pm \infty$.  Concretely, by embedding this space to a higher dimensional Euclidean space as fig.~\ref{GSWH}, one can visually understand the structure. Here embedding function is given as in the Lorentzian case, except for the sign of 5-dimensional metric: 
\begin{align}
l^2 (dz^2 + d a^2 + a^2 \Omega_{ij}dx^{i}dx^{j}) \rightarrow l^2 \left( (z'(t_{E})^2 + a'(t_{E})^2) dt_{E}^2 + a(t_{E})^2 \Omega_{ij}dx^{i}dx^{j} \right) = l^2 \left( N^2 dt_{E}^2 + a(t_{E})^2 \Omega_{ij}dx^{i}dx^{j}\right),
\end{align}
where the embedding function can be obtained as 
\begin{align}
 z(t_{E}) = \int^{t_{E}}_{0} d t_{E}\sqrt{N^2 - (\partial_{t_{E}}a)^2}.
\end{align}
Based on the standard prescription of WKB analysis, the tunneling probability from the flat space to the time crystal Universe can be estimated by the on-shell value of Euclidean action. We find that the probability is finite:
\begin{align}
S_{E} &=
6 l^2 M_{\mathrm{pl}}^2 \pi^2 \int_{-\infty}^{\infty} d t_{E} \left[
 \frac{1}{\cosh t_{E}}
+ \frac{\gamma}{4} \left(- \frac{1}{\cosh^2 \left(\frac{t_{E}}{2} \right)} + \frac{6}{\cosh^2 t_{E}} -  \frac{4}{\cosh^4 t_{E}}  \right) + {\cal O}(\gamma^2)
 \right] \notag\\
&= 6 l^2 M_{\mathrm{pl}}^2 \pi^3 \left(
 1 + \frac{2}{3 \pi} \gamma + {\cal O}(\gamma^2)
\right),
\end{align}
and hence our time crystal Universe can be nucleated from the flat space. This scenario is visualized as in fig.~\ref{AC1}.
Since two sides of the asymptotically flat region can be connected to Minkowski spacetimes, this instanton describes a tunneling process from $\mathrm{R}^3$ to $\mathrm{R}^3 \oplus S^3$, where $S^3$ is our time crystal Universe.
The tunneling probability of Giddings-Strominger is recovered by setting $\gamma = 0$.
Since the correction $\gamma$ is positive, nucleation probability of the time crystal Universe is slightly smaller than standard Giddings-Strominger's baby Universe.

\section{Tunneling from the time crystal to de Sitter Universe}
\label{sec4B}
In this section, we show, once the cosmological constant is tuned on, a time crystal Universe will transit to inflationary Universe by tunneling.

\begin{figure}[htbp]
\begin{minipage}{0.49\hsize}
\begin{center}
 \includegraphics[width=\hsize]{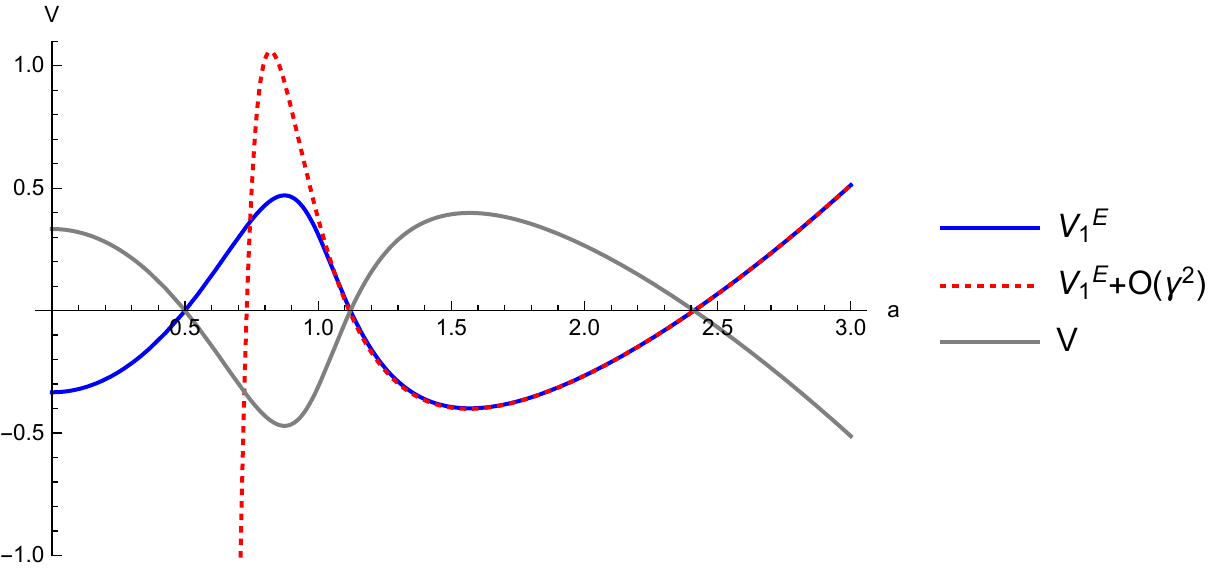}
 \caption{Euclidean potential $V^E$ in 2-form theory with a small cosmological constant ($\gamma = \frac{1}{4}$, $k = 1$,$\lambda = \frac{1}{2}$):
The gray plot is the Lorentzian potential and the red, dashed one is a Taylor expansion of Euclidean potential given by \eqref{VElambda}. 
Here the points of $V^{E} = 0$ are given as $a_{\mathrm{min}} = 0.5, a_{\mathrm{max}} = 1,12, a_{\Lambda} = 2.41$.}
\label{VE3}
\end{center}
\end{minipage}
\begin{minipage}{0.49\hsize}
\begin{center}
 \includegraphics[width=\hsize]{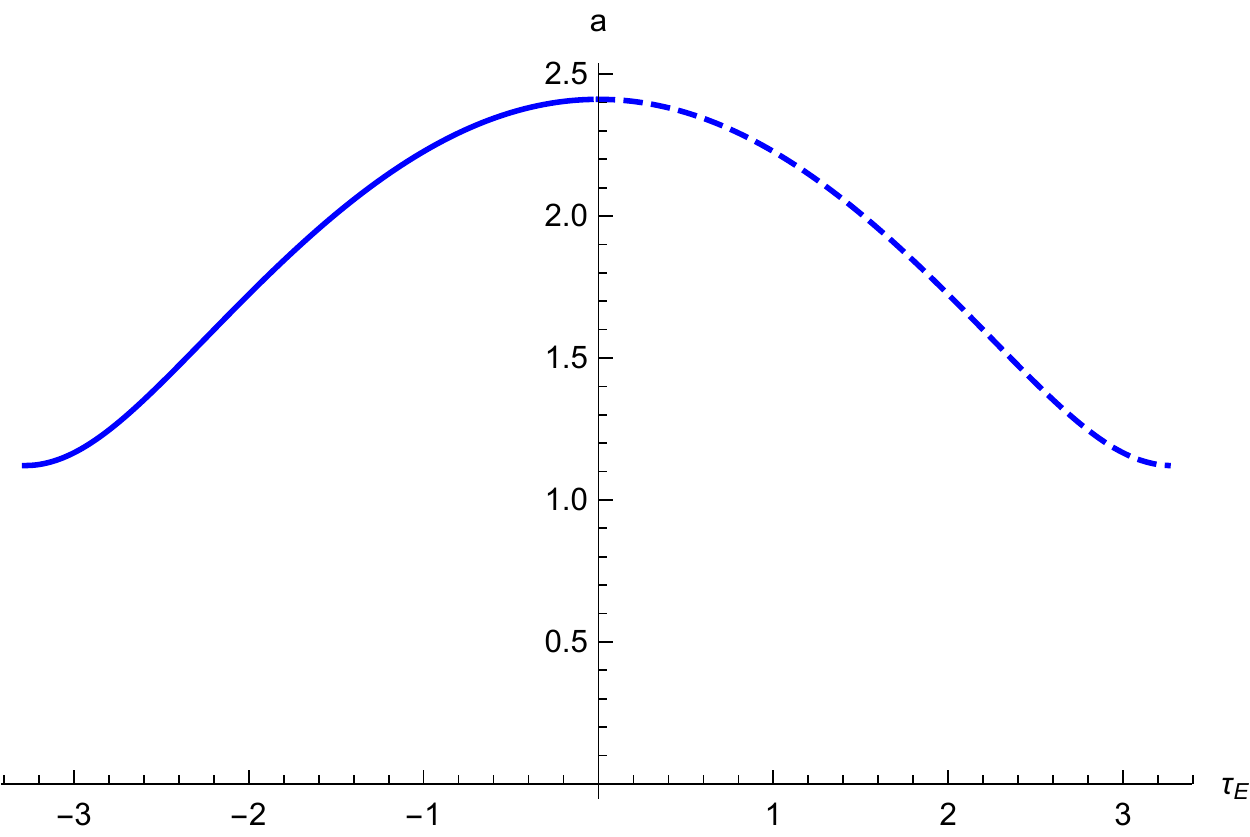}
 \caption{The scale factor $a$ as a function of the proper time $\tau_{E}$: Here only a period is plotted.}
\label{atauEplot}
\end{center}
\end{minipage}
\end{figure}

A potential with a positive small $\lambda$ is shown in fig.~\ref{VE3}. The tunneling from a time crystal Universe to an inflationary one will be described a bounce solution between $a_{\mathrm{max}}$ and $a_{\Lambda}$. As seen from fig.~\ref{VE3}, the Taylor expansion of the potential is good approximation and that is given by 
\begin{align}
 V_{E}(a;\gamma, \lambda, k = 1 ) = - 1 + \frac{1}{a^4} + \frac{1}{3} a^2 \lambda - \frac{1}{a^6} \left(
- 2 + \frac{1}{a^4} + \frac{1}{3} \lambda a^2
\right) \gamma + {\cal O}(\gamma^2).\label{VElambda}
\end{align}

As in the analysis of $\lambda = 0$ case, one can freely choose a time coordinate $t_{E}$ by fixing a functional form of $a(t_{E})$. However, we can not introduce the global time coordinate analytically, simply because it is difficult to derive the analytic expression for the points $a_{\mathrm{max}}$ and $a_{\Lambda}$. Instead, we will introduce a time coordinate with coordinate singularity.
The simplest choice is
\begin{align}
 a(t_{E}) = t_{E} \qquad t_{E} \in (a_{\mathrm{max}}, a_{\Lambda}),\label{aE2}
\end{align} 
and the lapse function $N$ can be obtained from the Friedmann equation as 
\begin{align}
 N(t_{E})^2 = \frac{1}{-V(t_{E})} = - \frac{t_{E}^4}{1 -t_{E}^4 + (\lambda/3) t_{E}^6}
- \frac{1 - 2 t_{E}^4 + (\lambda/3)t_{E}^6}{ t_{E}^2(1 - t_{E}^2 + (\lambda/3)t_{E}^6)^2} \gamma + {\cal O}(\gamma^2).\label{NE2}
\end{align}
There are coordinate singularities at $a = a_{\mathrm{max}}$ and $a_{\Lambda}$ where $N(t_{E}) \rightarrow \infty$ ($V(t_{E}) = 0$) and our time coordinate spans only a finite part of the whole space.
By integrating $N$, we can obtain the proper time $\tau$, 
\begin{align}
 \tau(t_{E})  = \int^{t_{E}}_{a_{\Lambda}} N(s) ds, \qquad t_{E} \in (a_{\mathrm{max}}, a_{\Lambda}) ,
\end{align}
where we set  $\tau = 0$ at the bounce point $a = a_{\Lambda}$.
Numerically scalar factor can be plotted as a function of the proper time $\tau_{E}$ as the solid curve in fig.~\ref{atauEplot}.  Similarly, the solution for $t_{E} > a_{\Lambda}$ can be obtained by assuming  
\begin{align}
 a(t_{E}) = 2 a_{\Lambda} - t_{E}, \qquad  t_{E} \in (a_{\Lambda}, 2 a_{\Lambda} - a_{\mathrm{max}}).
\end{align}
By repeating same procedure, we can  plot the scale factor $a$ as a function of $\tau_{E}$ for $(a_{\Lambda}, 2 a_{\Lambda} - a_{\mathrm{max}})$, which is the dashed curve in fig.~\ref{atauEplot}.

\begin{figure}[htbp]
\begin{minipage}{0.49\hsize}
\begin{center}
 \includegraphics[width=\hsize]{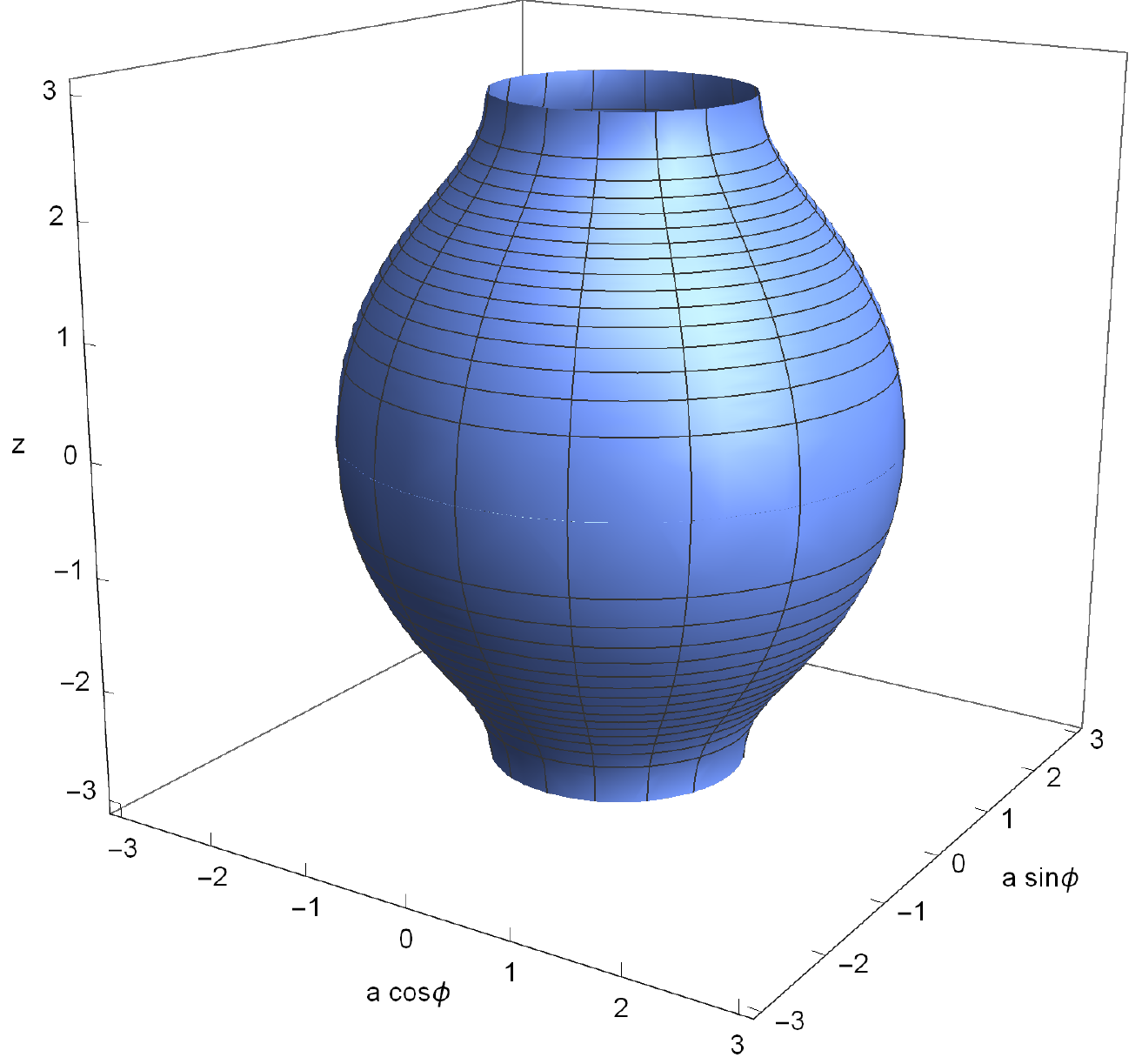}
 \caption{The instanton solution embedded in the flat Euclidean space}
\label{bounce.pdf}
\end{center}
\end{minipage}
\begin{minipage}{0.49\hsize}
\begin{center}
 \includegraphics[width=\hsize]{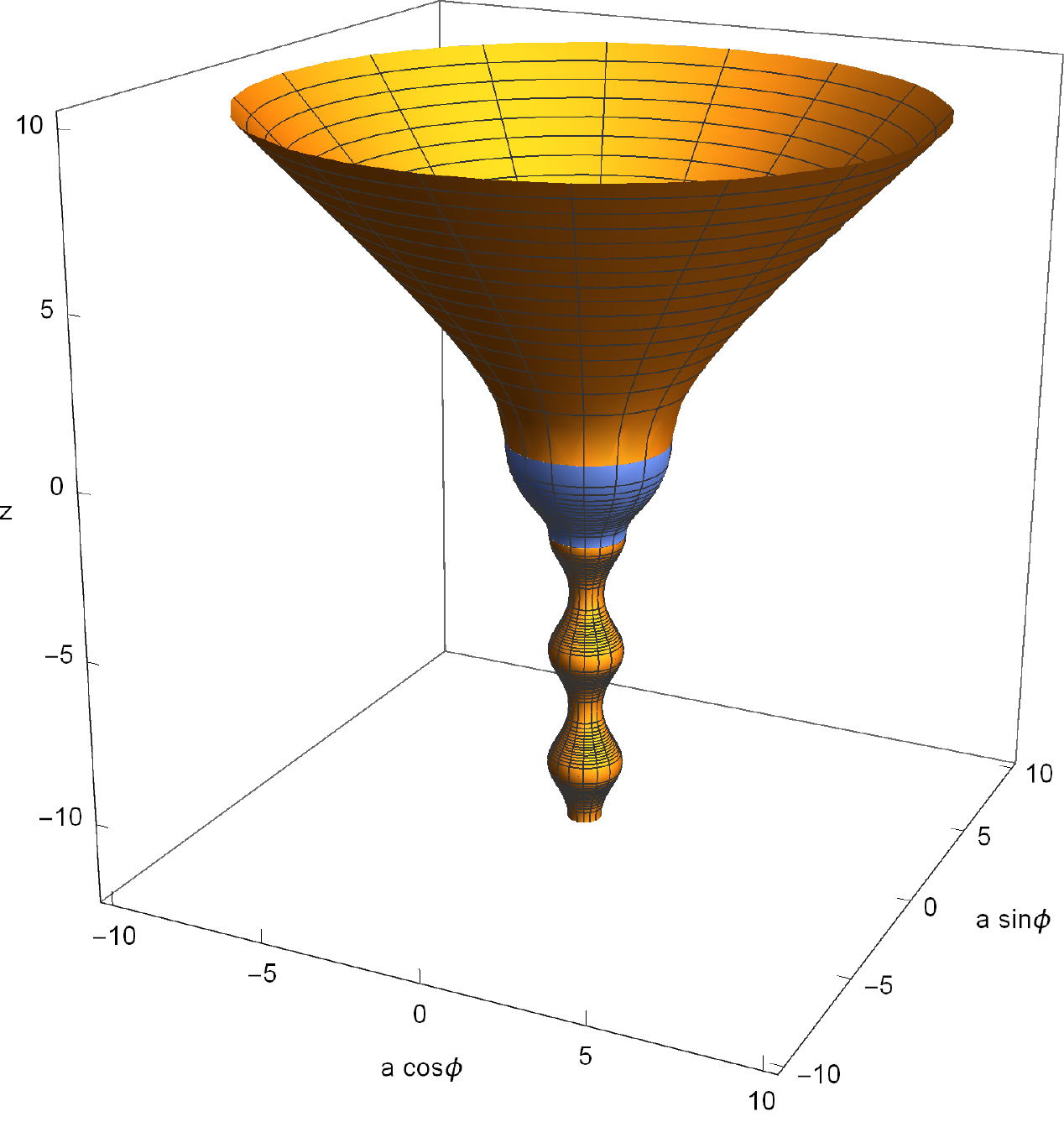}
 \caption{
Tunneling process from time a crystal Universe to de Sitter Universe mediated by our instanton solution:
The orange surfaces are embedded in Minkowski spacetime as well as the blue one is in Euclid space 
}
\label{TCUtodS}
\end{center}
\end{minipage}
\end{figure}

The embedding to the 5 dimensional Euclidean space can be obtained as in the $\lambda = 0$ case. Now embedding function $z$ is given by
\begin{align}
 z(t_{E}) = \int_{a_{\Lambda}}^{t_{E}} \sqrt{N(t_{E})^2 - 1}.
\end{align}
The embedding of the instanton solution is plotted in fig.~\ref{bounce.pdf}.

Finally we need to confirm that the transition probability is finite. Since we know the exact solution \eqref{aE2} and \eqref{NE2}, it is easy to estimate the value of the on-shell Euclidean action $S_{E}$ at least numerically.
The results of numerical calculations in the parameter region $\gamma \in (0.2, 0.8)$ and $\lambda \in (0.1, 0.8)$ can be fitted as
\begin{align}
 S_{E} \sim  6 \pi^2 l^2 M_{\mathrm{pl}}^2  \left( - \frac{ 4.0}{\lambda} + 3.4 + 1.1 \gamma \right).
\end{align}
For example, we obtain $S_{E} = - 4.4$ for the parameter $\gamma = \frac{1}{4}$ and $\lambda = \frac{1}{2}$. Since the value of $S_{E}$ is finite, we can conclude that our time crystal Universe can make a transition into an inflationary Universe through tunneling. 
The whole history of the tunneling process is described in fig.~\ref{TCUtodS}. There our instanton solution mediates a time crystal Universe and de Sitter Universe.
 
\section{Summary and Discussion}
We found that a 2-form gauge theory \eqref{SH}, as well as a dual Horndeski theory \eqref{Sphi}, has a solution describing a time crystal Universe, which is exactly periodic in time as shown in fig.~\ref{atauplot}.
By considering the semi-classical effect of quantum gravity, we found that (i) the time crystal Universe can be nucleated from the flat space when $\Lambda = 0$ and (ii) the time crystal Universe decays into de Sitter Universe when $\Lambda > 0$. The history of the
 transitions are visualized in fig.~\ref{AC1} and fig.~\ref{TCUtodS}.
 The last result suggest a new picture
  that the inflationary Universe is created from a time crystal by tunneling.  
Thus, our finding gives rise to a past completion of inflationary scenario.

Though our Euclidean solution provides a new interesting scenarios of early Universe, 
there are problems which need to be solved.  The first problem would be the stability of the time crystal Universe. Not only the time crystal Universe but any bouncing Universe in Horndeski theory tends to suffer from the stability issue~\cite{Kobayashi:2016xpl}. In case of the Universe with a spatial curvature, the stability is studied \cite{Akama:2018cqv} and found that the tensor perturbation is stable if and only if the following inequalities are satisfied:
\begin{align}
 {\cal F}_{T} : = M_{\mathrm{pl}}^2 - \beta X > 0, \qquad  {\cal G}_{T} : = M_{\mathrm{pl}}^2 + \beta X >0,
\end{align}
where positivity of ${\cal F}_{T}$ and ${\cal G}_{T}$ ensures the absence of gradient and ghost instability respectively. Thus the absolute value of the kinetic energy of scalar field $|X|$ must be finite. However one can check that $X$ goes to $- \infty$ as the scale factor $a$ approaches to its minimum $a_{\mathrm{min}}$. Thus near the lower turning point $a_{\mathrm{min}}$, tensor modes become ghosts. Since our action includes a negative mass dimension operator, it is natural to regard our theory as a low energy effective theory with a cut off scale $\Lambda_{\mathrm{cut}} < M_{\mathrm{pl}}$.
 Then, it is an interesting question that we can stabilize the cosmological perturbations by adding the higher dimensional operator like $K_{ij}K^{ij}$ without affecting the background dynamics as in the case of other time crystal solutions~\cite{Easson:2018qgr}.

Another problem is the Euclidean solution connecting to $a = 0$. From the plot of potential in fig.~ \ref{VE3}, one can find that the potential can be expanded as
\begin{align}
 V^{E}(a) = - \frac{1}{3} + {\cal O}(a^2).
\end{align}
Then we obtain the expression for the scale factor,
\begin{align}
 a(\tau) = \frac{\tau}{\sqrt{3}}  + {\cal O}(\tau^3),
\end{align}
and this leads to the curvature singularity at $\tau = 0$ because $R_{E} = 12/\tau^2 + {\cal O}(\tau^0)$. Though this does not leads to the divergence of the Einstein Hilbert term because  $ N a^3 R_E \propto - \tau$, the contribution from the 2-form kinetic term becomes negative infinity;
\begin{align}
 \frac{1}{12} \sqrt{g_{E}} {\cal G}_{E}{}^{\mu\nu\rho}{}_{\alpha\beta\gamma} H_{\mu\nu\rho}H^{\alpha\beta\gamma}= - \frac{9 \sqrt{3} l M_{\mathrm{pl}}^2}{2 \gamma \tau} + {\cal O}(\tau) \rightarrow - \infty.
\end{align}
Thus if we assume that the tunneling probability is given by $\mathrm{e}^{-S_{E}}$ it diverges and dominates over the other processes such as creation of the time crystal Universe. However it might be a matter of the prescription of quantum gravity. For example, it is well known that the sign in front of $S_{E}$ is flipped depending on the boundary condition of the wave function of the Universe (See, e.g. , Ref.\cite{Wiltshire:1995vk} for a review of quantum cosmology). Apart from this, we can simply say that solution near $a = 0$ is unreliable because a validity of effective field theory of our action should be broken near the curvature singularity.

A possibility of the presence of time crystal Universe before the inflationary phase provides interesting questions. For example, how can we know the signal of time crystal before tunneling process? We expect that the signal should be encoded by cosmological perturbations, if the problem of ghost instability is resolved. As a virtue of time crystal Universe fields on the time crystal Universe must be gradually amplified because of the Floquet's theorem, which is the time periodic version of the Bloch's theorem of usual crystals. Thus the cosmological perturbations are also expected to be amplified as the Universe oscillates.  In our scenario, time crystal Universe enters inflationary phase through tunneling effect. Then it is interesting to investigate how the perturbations around a tunneling background evolve. Are there remnants of amplified perturbations or are every perturbations washed out by tunneling process?  We will address this issue in future work.

\begin{acknowledgments}
  D.~Y. would like to thank Toshifumi Noumi and Masaru Siino for fruitful discussion. D.~Y. is supported by the JSPS Postdoctoral Fellowships No.201900294. J.~S. was in part supported by JSPS KAKENHI
Grant Numbers JP17H02894, JP17K18778, JP15H05895, JP17H06359, JP18H04589.
We are also supported by JSPS Bilateral Joint Research
Projects (JSPS-NRF collaboration) `` String Axion Cosmology.''
\end{acknowledgments}

\bibliography{ref}
\end{document}